\documentclass{article}
\usepackage[preprint]{colm2026_conference}

\usepackage{microtype}
\usepackage{hyperref}
\usepackage{url}
\usepackage{booktabs}
\usepackage{amsmath}
\usepackage{amssymb}
\usepackage{graphicx}
\usepackage{xcolor}
\usepackage{enumitem}
\usepackage{lineno}

\graphicspath{{../../figures/}}
\graphicspath{{./figures/}}

\definecolor{darkblue}{rgb}{0, 0, 0.5}
\hypersetup{colorlinks=true, citecolor=darkblue, linkcolor=darkblue, urlcolor=darkblue}

\newcommand{\KV}{\mathrm{KV}}
\newcommand{\kva}{\KV(B\!\mid\!A)}
\newcommand{\kvo}{\KV(B\!\mid\!\varnothing)}
\newcommand{\Dl}{\Delta}
\newcommand{\rd}{R(\delta)}

\title{Kamera: Unified Position-Invariant Multimodal KV Cache\\ for Training-Free Reuse}

\author{Bole Ma, Jan Eitzinger, Harald K\"ostler \& Gerhard Wellein \\
Erlangen National High Performance Computing Center (NHR@FAU) \\
Erlangen, Germany \\
\texttt{\{bole.ma,jan.eitzinger,harald.koestler,gerhard.wellein\}@fau.de}}

\begin{document}

\ifcolmsubmission
\linenumbers
\fi

\maketitle

\begin{abstract}
Multimodal agents repeatedly re-examine the same video frames, UI screenshots, and rendered artifacts as their context window slides and reasoning iterates, yet every look-back re-encodes from scratch, because prefix caches serve reuse only at a fixed leading position. We show this recompute is avoidable, and identify exactly what naive KV reuse loses: the cross-chunk conditioning a chunk absorbs from its neighbours.
This loss is asymmetric. The direct readout of a cached chunk is recovered exactly and for free by the standard state-merge. What remains is a diffuse, low-rank residue concentrated in deep layers, invisible to single-hop retrieval but precisely what multi-hop reasoning binds on. Blind reuse therefore leaves single-hop recall intact while halving multi-hop accuracy; this is the failure mode prior position-independent caches, designed for single-context or single-image reuse, do not address.
We repair it with a small, training-free low-rank conditioning patch stored alongside each position-free chunk. Reuse reduces to one operator across MLA, GQA, and MHA: exact RoPE re-rotation to any target position, plus the patch that restores cross-chunk binding. This makes three window operations cheap: \textbf{reorder} (one patch serves every ordering of a cached set), \textbf{sliding-window survival} (surviving chunks relocate via rotation only, zero re-encode), and \textbf{recall} (an evicted chunk is rehydrated by its patch, never re-encoded).
A rank-$m$ patch recovers full task accuracy on cross-chunk-binding benchmarks, MM-NIAH across two attention families and two-page doc-QA, at a fraction of the KV footprint, and reconstructs re-prefill KV to within bf16 rounding in a production SGLang kernel across six backbones. The conditioning signal is strongest in redundant vision and video streams, making our solution most impactful where multimodal agents spend their recompute budget.
\end{abstract}

\section{Introduction}

A multimodal agent's context routinely outgrows its attention window. A web agent slides a three-screenshot window over a monotonically growing transcript~\citep{webvoyager2024}; a long-video agent re-examines the same clip across many reasoning steps under changing prompts~\citep{lover12025,dvd2025}; a document agent traverses pages coarse-to-fine and re-accesses earlier ones after semantic filtering~\citep{docvstar2026}. In every case the same visual content is encoded, dropped, and \emph{seen again} at a new position, behind a changed prefix, or after eviction. Memory, not the nominal window, is the operative bound: even as context windows reach $1$M tokens, a locally served model holds only as much KV cache as device memory allows, so the sequence length used in practice is set by KV capacity, and the agent must continually slide, evict, and re-admit content. Managing context beyond the window is, operationally, managing this churn of reuse.

\begin{figure}[t]
\centering
\includegraphics[width=\linewidth]{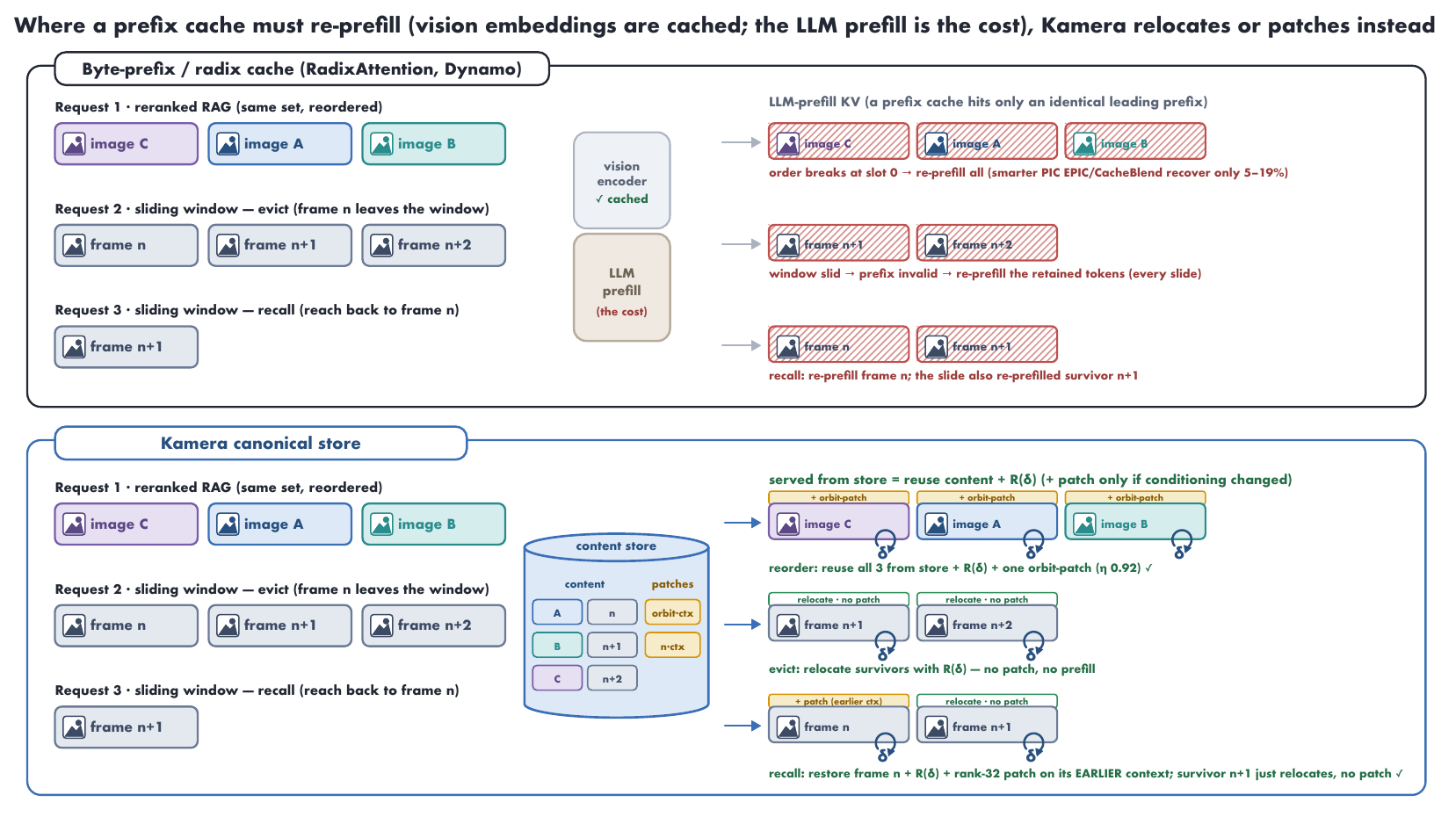}
\caption{Three reuse patterns beyond the window, and what each costs. \textbf{Top (prefix/radix):} cached vision embeddings still miss whenever reuse sits at a shifted position (reorder, slide, look-back), so the LLM prefill re-runs. \textbf{Bottom (Kamera):} chunks stored position-free; reorder reuses one orbit-patch, a survivor relocates for free ($\rd$, no patch), and a recalled chunk is rehydrated by a patch on its fixed earlier context. Survivors cost a rotation; only recall costs a patch.}
\label{fig:cases}
\end{figure}

Reuse is enormously cheap when it works: encoding a $1024$-token video segment costs $\approx\!230$\,ms of vision-tower compute, while replaying its stored KV costs $\approx\!5$\,ms~\citep{zheng2024sglang,kwon2023vllm}. But production caches express only one shape of reuse, because they treat the KV store as a \emph{position-indexed} structure. A prefix cache is an \emph{array}: a contiguous span addressed by absolute position, so evicting the oldest token shifts every position behind it, an $O(n)$ re-prefill. A radix cache adds a \emph{tree} of prefixes shared across requests~\citep{zheng2024sglang}. Both reuse a chunk \emph{only} while it sits at a fixed leading position behind a byte-identical prefix; the moment the window slides, the prefix changes, or the chunk is recalled at a new offset, the cache \emph{misses} and the engine re-encodes and re-prefills from scratch (Fig.~\ref{fig:cases}, top). The recurring multimodal patterns above (sliding windows, reorderings, look-backs) are radix misses by construction. Making a chunk \emph{position-free} changes the structure available: the same store can act as a \emph{deque}, evicting and admitting at either end of the window in $O(1)$, and, keyed by content rather than offset, points toward a content-addressed \emph{hash table} of reusable chunks.

\paragraph{Why the miss is not fundamental.} Re-prefilling a chunk at a new position recomputes two things that need not be recomputed. First, \emph{position}: a chunk's keys differ across offsets only by a RoPE phase rotation, which composes exactly ($R(\delta)R(p){=}R(p{+}\delta)$), so relocation is an algebraic re-rotation, not a forward pass. Second, \emph{conditioning}: prefilling a chunk $B$ (the content we cache and reuse) after an antecedent $A$ (whatever context precedes it in the window) lets $B$'s tokens absorb $A$ (coreferences resolved, entities bound). Concatenating independently cached chunks loses this cross-chunk conditioning, and \emph{only} this, because the other cross-attention (readout, what a query reads out of a chunk) is recovered exactly by the log-sum-exp state-merge that FlashAttention and ring/star attention already perform~\citep{dao2022flashattention}. Reuse is thus lossless for single-hop questions and silently breaks multi-hop ones: on a two-page document task, single-hop accuracy is unchanged under reuse ($0.57$) while multi-hop accuracy falls $0.41\!\to\!0.28$ (MLA) and $0.28\!\to\!0.15$ (GQA). The model still answers fluently; it just stops resolving ``the object shown earlier.''

\paragraph{What we restore, and how.} We name the lost term, $\Dl=\kva-\kvo$ ($B$'s key/value with $A$ in front of it minus $B$ cached alone, $\varnothing$ marking the absent antecedent; \S\ref{sec:bg}), measure its shape, and attack it with the operator its shape dictates. The deficit is \emph{diffuse across tokens} (no small ``important-token'' set; an oracle token selector needs $\approx\!50\%$ of tokens), yet \emph{low-rank in features} ($\approx\!90\%$ of its output-relevant energy in $\approx\!32$ directions) and \emph{deep} (negligible in shallow layers). So the prevailing fix, recomputing a few important tokens (CacheBlend~\citep{yao2024cacheblend}, VLCache~\citep{qin2025vlcache}, EPIC~\citep{hu2024epic}, MPIC~\citep{zhao2025mpic}), corrects the wrong axis. These methods were validated on \emph{single-context} compression or \emph{single-image} recurrence under prompt staleness, where the reused KV is nearly valid and a few token recomputes suffice; \emph{none target the cross-chunk binding} a windowed agent breaks, and on a real multi-hop video task they recover only a fraction of the answer flips (mean ${\approx}20\%$, $\le\!36\%$ at any token budget; against the patch's $97\%$; \S\ref{sec:cost}). We instead store each chunk as a \emph{position-free canonical} $\kvo$ plus a rank-$m$ \emph{conditioning patch}, and reuse it with
\begin{equation}
\widehat{\KV}(B\!\mid\!A)\;=\;\underbrace{\rd\cdot\kvo}_{\text{relocate (exact)}}\;+\;\underbrace{U_m V_m^{\!\top}}_{\text{rank-}m\text{ patch (conditioning)}}
\label{eq:operator}
\end{equation}
where $\rd$ re-rotates the stored keys' RoPE phase to the new position and $U_mV_m^{\!\top}$ is the top-$m$ SVD of $\Dl$, supervised by a single conditioned forward at compile time. The same operator covers MLA~\citep{deepseekv2}, GQA~\citep{ainslie2023gqa}, and MHA once each layout is read through a single \emph{content $\mid$ rope} split (\S\ref{sec:operator}). It is training-free and runs recompute-free inside a production engine.

\paragraph{The window operations this buys.} Separating position from conditioning, and storing the canonical apart from the patch, turns three window operations from re-prefills into millisecond cache edits (Fig.~\ref{fig:cases}, bottom; \S\ref{sec:window}). (i)~\textbf{Reorder} (reranked RAG, reshuffled frames, multi-image VQA): one stored \emph{orbit} patch serves every ordering of a cached set (verified exhaustively at $K{=}3$). (ii)~\textbf{Sliding-window survival}: when the window slides and an older chunk leaves, the chunks that \emph{stay} need only the exact re-rotation, no patch, and stay near-lossless. (iii)~\textbf{Recall} (reversible eviction): an evicted chunk's conditioned KV can be dropped while its canonical is kept (or recomputed for $\approx\!1/8$ the bytes from a standard vision-embedding cache), and re-instated later at \emph{any} position by a fresh patch on its now-fixed earlier context, with no vision re-encode. We measure all three across GQA, deepstack-GQA, and MLA, and find a clean asymmetry: \emph{recall costs a patch; survivors cost only $\rd$}.

\paragraph{Contributions.}
\begin{itemize}[leftmargin=1.2em,itemsep=1pt,topsep=2pt]
\item A \textbf{position/conditioning separation} that makes a cached multimodal chunk position-free: an exact RoPE relocation plus a rank-$m$ conditioning patch, one training-free operator across MLA, GQA, and MHA (\S\ref{sec:operator}). It is grounded in a diagnosis of what reuse drops, the cross-chunk conditioning that halves accuracy on tasks needing cross-chunk binding while leaving single-hop readout intact, which we measure to be diffuse-in-tokens yet low-rank-in-features and deep across six backbones (\S\ref{sec:shape}). Because the deficit is deep, a \emph{non-universal} cheaper variant patches only the deep layers at roughly half the bytes, with the depth budget model-dependent.
\item A \textbf{measured account of three window operations} prefix caching cannot serve (reorder, sliding-window survival, and recall under reversible eviction), including the eviction asymmetry (recall needs a patch; survivors need only relocation) across three attention families (\S\ref{sec:window}).
\item \textbf{Recompute-free serving} (amortized after ${\approx}9$ reuses) in SGLang's production paged-attention kernel and KV pool, reconstructing the re-prefill KV to within bf16 rounding (residual next-token KL $\approx\!10^{-3}$, two orders below blind reuse) with downstream accuracy matching the ceiling, and the cost win on the memory axis (full accuracy at a small fraction of the KV bytes), bounded to the redundant-stream regime where the effect lives (\S\ref{sec:cost}).
\end{itemize}

\section{What reuse loses: conditioning, not readout}
\label{sec:bg}

When a decoder attends over $\KV(A)\,\Vert\,\KV(B)$, two mechanisms are in play. \textbf{Readout} is the value a query pulls out: attention over the union of two key sets equals attending each separately and merging by softmax mass, $o=(1-\mu)\,o_B+\mu\,o_A$, the log-sum-exp state merge already used by FlashAttention and ring/star attention~\citep{dao2022flashattention}. A query reading an answer out of one chunk does not care that the chunk was cached separately, so \emph{single-hop reuse is exactly lossless}. \textbf{Conditioning} is what $B$'s own key/value vectors encode. If $B$ is prefilled alone its KV is $\kvo$; if prefilled after $A$ its tokens absorb $A$, giving $\kva$. The only quantity reuse loses is the deficit
\begin{equation}
\Dl \;=\; \kva \;-\; \kvo .
\label{eq:delta}
\end{equation}
A 4D-attention-mask oracle that blocks $B\!\not\to\!A$ in a single forward reproduces the loss at $B$'s exact positions: the failure is a binding deficit written into the KV, not a boundary attention artifact, so sink/boundary fixes (EPIC-style) cannot repair it. This is the term Eq.~\ref{eq:operator}'s patch supplies.

\section{The operator: relocate exactly, patch the conditioning}
\label{sec:operator}

\begin{figure}[t]
\centering
\includegraphics[width=\linewidth]{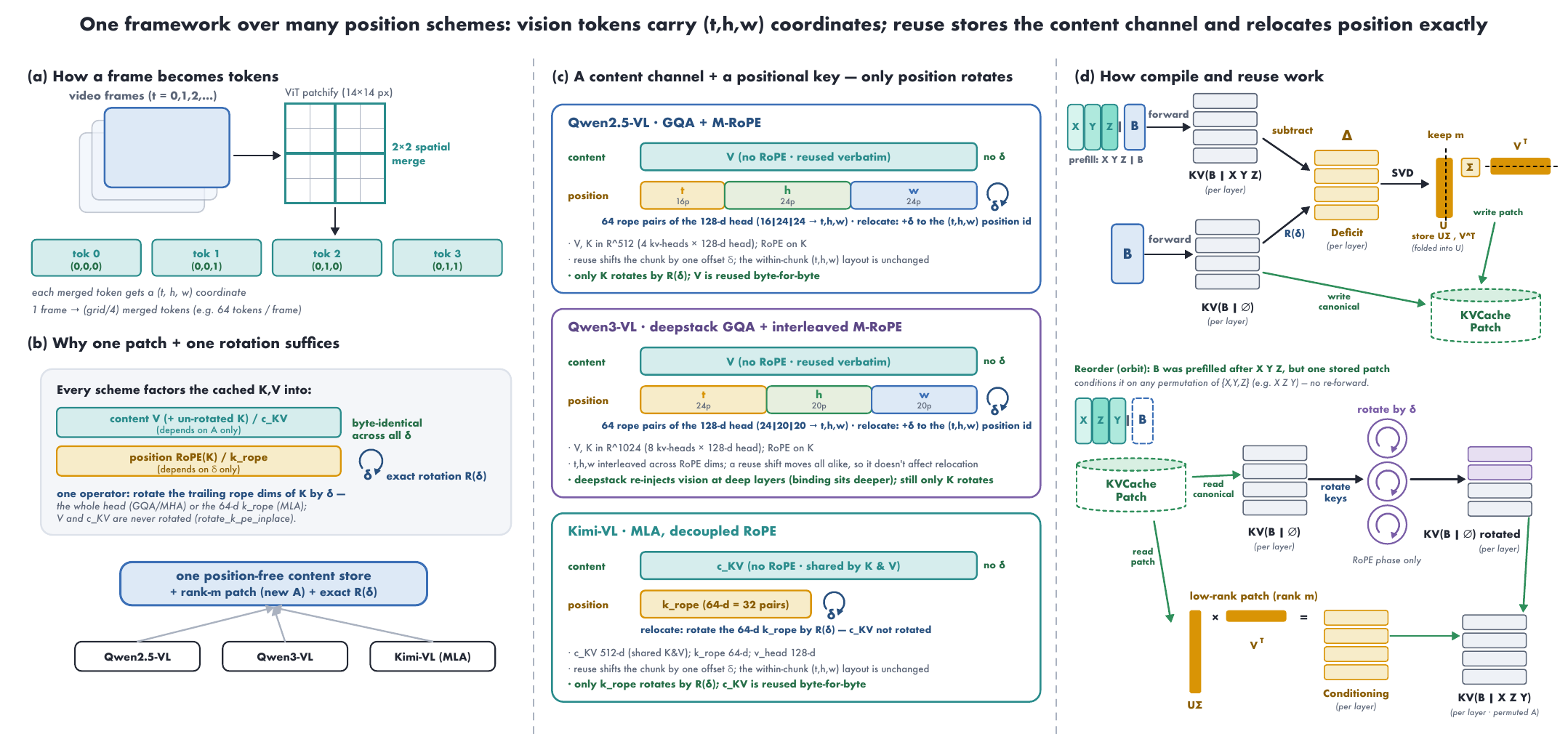}
\caption{Position-invariant storage across attention families. \textbf{(a)} the vision tower splits a frame into tokens, each tagged with a time/height/width coordinate. \textbf{(b,c)} every backbone splits its cached keys/values into a position-free \emph{content} part (the MLA latent, or the GQA/MHA value) and a \emph{positional} part (the rotary phase on the key); reuse re-rotates only the positional part to the new location---advancing all three coordinates together, so the blocked vs.\ interleaved layout does not matter---and reuses the content part byte-for-byte. One mechanism for Qwen2.5-VL (GQA), Qwen3-VL (deepstack), and Kimi-VL (MLA). \textbf{(d)} compile vs.\ reuse: one conditioned forward measures the deficit (what the chunk would have absorbed from its antecedent); its few dominant directions are stored alongside the content, and each later request re-rotates the keys and adds the patch back with no forward. One stored patch reconditions any ordering of the cached set.}
\label{fig:unify}
\end{figure}

Eq.~\ref{eq:operator} has two parts that answer to \emph{different} variables, which is what makes the cache position-free (Fig.~\ref{fig:unify}). \emph{Relocate}: re-rotate $B$'s keys by $\delta=p_1-p_0$. Because RoPE~\citep{su2024roformer} composes, $R(\delta)R(p_0)=R(p_1)$ exactly; $V$ is untouched. This term depends on the offset $\delta$ alone. \emph{Patch}: add $U_mV_m^{\!\top}$, the rank-$m$ correction supplying the binding $B$ would have absorbed from $A$. This term depends on the antecedent $A$'s \emph{content} alone, not on $\delta$. Hence relocating $B$ at fixed $A$ is absorbed \emph{exactly} by $\rd$ (the stored content channel is byte-identical across positions, so the same patch transfers unchanged, the reuse primitive), while changing the antecedent forces a new patch (conditioning $B$ on $A$ versus on neutral filler at the same position leaves the full deficit, so the patch encodes \emph{which} $A$).

\paragraph{One mechanism for MLA, GQA, and MHA.} These three families span the KV-sharing axis, from MLA's compressed latent through GQA's grouped heads to MHA's full per-head keys, yet collapse to one pipeline once each is read as a \emph{content} channel (position-free, what we store and patch) plus a \emph{RoPE} channel (what we rotate), Fig.~\ref{fig:unify}. \textbf{MLA} is the cleanest \emph{positionally}: the latent $c_{KV}$ carries no RoPE, so relocation only re-rotates the $64$-dim decoupled $k_{pe}$. The \emph{conditioning} patch touches both channels: the latent alone leaves a residual ($\approx\!8\times$ the floor), closed by a small added $k_{pe}$-band patch (content goes most of the way, the addressing band needs the rest); MLA then recovers comparably to GQA/MHA (\S\ref{sec:cost}). \textbf{GQA} has no separate content channel, so we relocate the full key by re-applying RoPE and patch both $K$ and $V$ per KV-head. \textbf{MHA} is GQA with one KV head per query head, treated identically. ``Split content $\mid$ RoPE; store the content channel; at reuse, rotate RoPE and patch content'' is the same pipeline in all three.

\paragraph{Forming and applying the patch.} The patch is supervised by \emph{one} forward, paid once and amortized. At compile time we run a single conditioned forward over $[\,\text{prefix}\cdot A\cdot B\,]$, read $B$'s conditioned KV $\kva$, subtract the stored relocated $\rd\cdot\kvo$ to obtain $\Dl$, and keep its top-$m$ SVD factors $\{U_m,V_m\}$ ($\approx\!2\%$ of the page). At serve time we apply Eq.~\ref{eq:operator} with \emph{zero} forwards: a per-layer RoPE rotation plus a GEMM into the paged KV cache, bandwidth-bound and needing no kernel surgery beyond a cache hook (listings in App.~\ref{app:impl}). Re-prefill pays a forward on \emph{every} request; we pay it once at compile and every reuse thereafter is forward-free. The win is therefore amortized: it materializes once the same chunk recurs (break-even $\approx\!9$ reuses against a prefill-per-reuse baseline, \S\ref{sec:cost}), the concentrated-reuse regime a long-horizon multimodal agent generates.

\section{The shape of the lost term dictates a feature patch}
\label{sec:shape}

\begin{figure}[t]
\centering
\includegraphics[width=\linewidth]{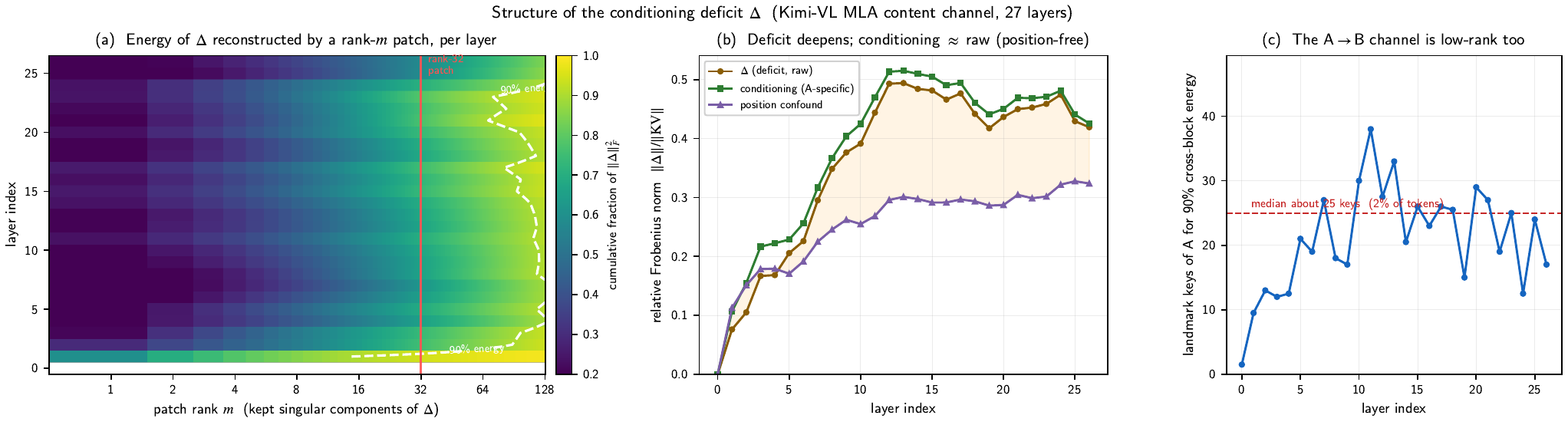}
\caption{Structure of the conditioning deficit (Kimi-VL, MLA content channel, $27$ layers). \textbf{(a)} how much of each layer's deficit a rank-$m$ patch captures: the useful knee is near rank $32$, well left of where $90\%$ of the raw energy sits, so the patch keeps the output-relevant directions rather than all the energy. \textbf{(b)} the deficit grows with depth and is almost entirely \emph{conditioning}, not position, so the patch corrects content. \textbf{(c)} the link from antecedent to chunk is itself low-rank (about $25$ landmark keys carry $90\%$). Low-rank and deep, not token-sparse---which is why a feature patch beats recomputing tokens.}
\label{fig:shape}
\end{figure}

If $\Dl$ were large and unstructured nothing cheap could help. It is highly structured along three axes, and the structure decides the design (Fig.~\ref{fig:shape}).

\textbf{Low-rank in features.} Stacking $\Dl$ over $B$'s tokens, the functional rank that recovers the output distribution is $m\!\approx\!32$ (the KL plateau), far below the $\sim\!120$ components holding $90\%$ of $\Dl$'s raw energy. The patch needs only the top, output-relevant directions. Sweeping $m$, conditioning-KL knees at $m\!\approx\!8$--$16$ and plateaus by $32$ on \emph{every} structure (GQA-512, GQA-1024, MoE, MLA); the \emph{saturating} rank is absolute, not a width fraction. The directions are moreover \emph{shared across items}: a fixed per-layer basis pooled over $(A,B)$ pairs recovers a held-out deficit as well as that item's own SVD, and transfers across content/task, so the patch's directions are a property of the model and only the coefficients are item-specific.

\textbf{Diffuse across tokens.} Low-rank in features does not mean sparse in tokens. There is no small binding-token set: an oracle that selects tokens by true $\Dl$-magnitude needs $p\!\approx\!0.5$ to recover most of the gap, and a first-$k$ ``carve'' is worse than nothing. The few binding directions touch a little of \emph{most} tokens, so token-recompute methods aim at the wrong axis.

\textbf{Deep.} $\Dl$'s relative norm grows with depth ($0.08\!\to\!0.49$, shallow$\to$deep). A single-layer injection explains $\approx\!27\%$ of the final deficit applied shallow but $\approx\!97\%$ applied deep, with no shallow shortcut. Together: a thin patch \emph{can} carry the loss (low-rank), a token subset \emph{cannot} (diffuse), and the correction must live \emph{deep}.

\section{Reuse beyond the window}
\label{sec:window}

The separation of \S\ref{sec:operator} turns three patterns a windowed agent generates, each a prefix-cache miss, into cheap cache edits. We probe all three on cached video segments across GQA (Qwen2.5-VL~\citep{bai2025qwen25vl}), deepstack-GQA (Qwen3-VL~\citep{bai2025qwen3vl}), and MLA (Kimi-VL~\citep{kimivl2025}), reporting $\eta$ (the fraction of the blind-reuse$\to$re-prefill KL gap an arm closes) and, where decisions matter, flip-recover (recovery on the subset where blind reuse flips the re-prefill answer).

\paragraph{Reorder: one orbit-patch for all orderings.} The cleanest miss is identical chunks in a different order (multi-image VQA, reranked RAG, reshuffled frames), where content is byte-identical and only positions and cross-chunk conditioning change. Permuting the predecessor set and comparing the stored canonical-order patch (\emph{transfer}), the ordering's own patch (\emph{exact}), and a single patch averaged over the permutation orbit with the test ordering held out (\emph{orbit}): the orbit patch is near-exact, $\eta_{\text{orbit}}{=}0.92\!\approx\!\eta_{\text{exact}}{=}0.94$ on Qwen2.5-VL, and architecture-universal (deepstack $0.87\!\approx\!0.92$, MLA $0.87\!\approx\!0.94$, MHA DeepSeek-VL~\citep{lu2024deepseekvl} $0.89\!\approx\!0.93$). The raw deficit is \emph{not} order-invariant ($\lVert\Dl_\pi-\Dl_{\text{id}}\rVert/\lVert\Dl_{\text{id}}\rVert{=}0.43$--$0.53$), yet the orbit mean captures the recoverable component \emph{without degrading as the orbit grows}: it tracks the per-ordering exact patch through $K{=}6$ --- tested \emph{exhaustively} over all orderings at $K{=}3$ and $K{=}4$ ($3!$ and $4!$), then sampled at $K{=}6$ ($24$ of $720$); $\eta_{\text{orbit}}{=}0.92/0.93/0.94$ at $K{=}3/4/6$ on Qwen2.5-VL, $0.87/0.85/0.83$ on MLA. So one orbit-patch serves every ordering of the set.

\paragraph{Sliding-window survival: relocate for free.} When the window slides and the oldest chunk leaves, the chunks that \emph{remain} shift to new positions but keep their original antecedents. We evict the leading chunk and ask what the survivors need. The answer is: only the re-rotation. Keeping a survivor's conditioned KV as-is and applying $\rd$ is near-lossless on GQA and MLA (keep-as-is KL $0.015$/$0.023$), because the evicted chunk's already-absorbed influence is small next to the surviving conditioning. The deepstack backbone is the exception (keep-as-is KL $0.113$, $5$--$7\times$ higher): its deep visual re-injection makes even a survivor sensitive to the evicted antecedent. Where it bites, the \emph{removal} deficit is the low-rank deep dual of the addition deficit (deep rel-norm $\approx\!3\times$ shallow; $90\%$-energy rank $36$--$44$), so a rank-$64$ removal patch recovers it ($\eta{=}0.82$--$0.87$). The practical rule: \emph{slide the window for free; patch the deepstack survivor if you need exactness} (Table~\ref{tab:evict}).

\paragraph{Recall: reversible eviction patches the fixed past.} A windowed agent must sometimes reach back to a chunk it evicted. Because the canonical is what we store and the patch is what we add at reuse, eviction is \emph{reversible}: drop the conditioned KV, keep the canonical, and re-instate later at any position. The question is whether the \emph{stored} patch can be replayed, and it cannot. A patch frozen at eviction goes \emph{stale} as the window turns over, decaying monotonically from $\eta\!\approx\!0.9$ at no turnover to actively harmful at full turnover ($\eta{=}-0.68$ GQA, $-2.85$ deepstack; MLA decays more gently to $+0.29$), recovering $0$--$25\%$ of answer flips. A \emph{fresh} rank-$32$ patch, conditioned on the chunk's now-fixed \emph{earlier} context (hence itself storable and never stale), restores rebuild quality ($\eta{=}0.87$/$0.96$/$0.81$; flip-recover $0.75$/$1.0$/$0.67$ vs.\ full re-prefill's $1.0$). So storing the clean chunk is necessary but not sufficient: \emph{recall costs one patch}, formed on the stable past, and the vision encoder never re-runs. This is what heuristic single-context eviction~\citep{zhang2023h2o,wan2024lookm} cannot do: it discards position-baked KV that cannot be re-placed at a new offset, so a look-back today pays a full re-encode.

\begin{table}[t]
\centering
\small
\begin{tabular}{l l c c c}
\toprule
event & arch & stale / keep-as-is & fresh patch $\eta$ & flip-recover (stale$\to$patch) \\
\midrule
recall          & GQA       & $\eta{=}-0.68$ & $0.87$ (r$32$) & $0.25\to0.75$ \\
(full turnover) & deepstack & $\eta{=}-2.85$ & $0.96$ (r$32$) & $0.00\to1.00$ \\
                & MLA       & $\eta{=}+0.29$ & $0.81$ (r$32$) & $0.00\to0.67$ \\
\midrule
survivor        & GQA       & $\mathrm{KL}{=}0.015$ & $0.87$ (r$64$) & --- \\
(head evicted)  & deepstack & $\mathrm{KL}{=}0.113$ & $0.83$ (r$64$) & --- \\
                & MLA       & $\mathrm{KL}{=}0.023$ & $0.82$ (r$64$) & --- \\
\bottomrule
\end{tabular}
\caption{Eviction is asymmetric. \emph{Recall} needs a fresh patch: the stale stored patch turns harmful as the window turns over ($\eta{<}0$ on both GQA backbones), while a rank-$32$ patch on the chunk's fixed earlier context tracks full rebuild. \emph{Survivors} need only $\rd$: keeping their KV as-is is near-lossless on GQA/MLA; only the deepstack backbone leaves a (low-rank) removal deficit. Cached video segments, $n{=}25$--$32$/model; probe details in App.~\ref{app:menu}.}
\label{tab:evict}
\end{table}

\paragraph{The window mechanics, end to end.} A single slide composes the three: as the oldest frame leaves, survivors re-rotate by the slide offset (free) and a later recall rehydrates the dropped frame from the canonical store with a patch on its preceding context. The orchestrator can evict aggressively, since a mis-eviction costs a cheap rehydrate, not a re-encode. Storing the canonical apart from the patch makes conditioning a reversible \emph{switch} (free clean-view overwrite, rank-$m$ re-add), enabling context-clean forks and a near-free disposability test (App.~\ref{app:ctx}).

\section{Fidelity, deployment, and cost}
\label{sec:cost}

\paragraph{The feature patch reaches the re-prefill ceiling; the token axis does not.} Against the audit-faithful named PIC baselines given the \emph{same} relocated KV (token baselines recompute their selected tokens \emph{in context}, the strongest CacheBlend form), the rank-$m$ feature patch closes $98$--$100\%$ of the reuse$\to$re-prefill KL across MLA/GQA, Dense/MoE while token-recompute closes $10$--$71\%$, and on the pooled items where reuse flips the answer the patch restores the re-prefill decision $96\%$ of the time versus $21$--$44\%$ for token baselines (Fig.~\ref{fig:baselines}). The accuracy gap is real where the answer needs cross-chunk binding: on MM-NIAH~\citep{wang2024mmniah}, blind KV reuse cuts Qwen2.5-VL accuracy roughly in half --- both on retrieval-image, a single needle that must bind to the query ($0.74\!\to\!0.38$), and on the multi-hop reasoning-image split ($0.59\!\to\!0.41$) --- while a rank-16 patch restores the ceiling ($0.72$ / $0.64$); the same gap and recovery hold on a second KV family (Kimi-VL, MLA), the patch beating the token baselines on both (Table~\ref{tab:mmniah}). On MileBench's~\citep{song2024milebench} temporal suite (cross-frame binding over a cached clip), the rank-64 patch recovers the re-prefill decision $97\%$ of the time while the named token baselines at a $10$--$15\%$ budget (VLCache, CacheBlend) stay near the blind floor (App.~\ref{app:menu}, Table~\ref{tab:milebenchT}); only a shallow partial \emph{re-prefill} keeps up, confirming the conditioning is born deep. At a matched KV-byte budget the patch exceeds every token-axis recovery (Table~\ref{tab:matched}): decisively over first-$k$ and a ShadowKV-style low-rank-$K$ reconstruction (which rebuilds absolute $K$, which the canonical already has, not the conditioning delta), and significantly though modestly over an oracle query-aware token selector, so the axis is wrong, not just the selector.

\paragraph{Recompute-free on a live engine.} In SGLang's production FlashAttention-3~\citep{shah2024fa3} paged-attention kernel and KV pool (radix disabled), relocating a cached segment to a $\delta\!\neq\!0$ mid-sequence position the prefix-keyed radix \emph{scheduler} cannot express, the reconstructed KV writes into the pool within one bf16 ULP of recompute, and the resulting next-token KL sits at $\approx\!10^{-3}$ across four backbones, two orders below blind reuse ($0.03$--$0.12$). Downstream, recompute-free splice$+$patch tracks the re-prefill answer $89$--$95\%$ of the time and matches its \emph{accuracy} ceiling within $1$--$3$ points on Video-MME~\citep{fu2024videomme} and EgoSchema~\citep{mangalam2023egoschema} (Qwen2.5-VL, Kimi-VL; higher-rank patches reach $97$--$100\%$ per-item agreement, App.~\ref{app:obs:deploy}).

\paragraph{The cost is on memory.} Decode is memory-bandwidth-bound (its arithmetic intensity sits far below the H100 compute ridge), so the binding resource is KV bytes. A rank-64 patch matches full multi-hop accuracy at $\approx\!25\%$ of the segment's KV bytes (rank-16: $\approx\!6\%$), a fraction that holds on \emph{either} layout --- the per-head $K/V$ page or the MLA latent ($c_{KV}$ plus $k_{pe}$) page --- since the patch and the page scale together. The forming forward amortizes after $\approx\!9$ reuses against a prefill-per-reuse baseline (near-immediate against full recompute), and replacing the per-reuse LLM prefill with the forward-free patch-apply yields up to $29\times$ TTFT on long video (prefill-only; larger against full recompute, which also re-runs the vision encoder). The capacity sharing, amortization, and full TTFT decomposition are in App.~\ref{app:obs:deploy}.

\paragraph{Scope.} The effect is mechanism-bounded. Vision and video show the gap and recover, audio a smaller, partly-recovered gap (Qwen2.5-Omni, $n{=}40$); in our 2-chunk text setup, dense text (MuSiQue 2-hop~\citep{trivedi2022musique}, the two supporting paragraphs) shows \emph{no} gap (blind $\approx$ re-prefill), since the loss is a property of \emph{redundant} token streams whose meaning lives in cross-chunk binding (App.~\ref{app:obs:scope}). The open problem is cheap estimation of $\Dl$ without the conditioned forward, which we bound with clean negatives in App.~\ref{app:menu}.

\section{Related work}
\label{sec:related}

\paragraph{Context management beyond the window.} A growing line takes multimodal KV reuse seriously but pushes an \emph{orthogonal} axis: compressing or evicting \emph{within one growing context}. Streaming-video and driving models evict or compress KV memories (StreamingVLM~\citep{xu2025streamingvlm}, StreamMem~\citep{streammem2025}, HERMES~\citep{hermes2026}); robotics VLAs cache static frame tokens across control steps (VLA-Cache~\citep{vlacache2025}); agent frameworks add explicit look-back retrieval (PAL-UI~\citep{palui2025}, Embodied VideoAgent~\citep{embodiedvideoagent2025}). That several must \emph{selectively retain} relevant past KV rather than drop it freely corroborates that past context is load-bearing, the conditioning we name. But almost all operate within a single growing stream where recency suffices and reuse sits at stable positions; they do not re-prefill across requests. Our value is the complementary regime they leave open (cross-request, cross-position, multi-hop reuse, the prefix-cache miss), where agents pay: one screenshot sent to a planner and a grounder~\citep{seeact2024}, a page re-prefilled under each query's neighbours~\citep{m3docrag2024}, a clip re-examined per reasoning step. Where the image stays at a \emph{fixed} prefix, prefix caching already serves it and we add nothing.

\paragraph{Position-independent caching repairs the wrong axis.} The closest line reuses non-prefix KV and repairs the cross-chunk loss by \emph{selective token recompute}: CacheBlend~\citep{yao2024cacheblend}, CacheClip~\citep{cacheclip2025}, KEEP~\citep{keep2026}, KVLink~\citep{kvlink2025}, EPIC~\citep{hu2024epic}, MPIC~\citep{zhao2025mpic}, VLCache~\citep{qin2025vlcache}, all assuming the loss is token-sparse; our diffuse-token diagnosis shows that premise does not transfer to cross-chunk conditioning (in-context token recompute is a Pareto-worse axis here, reaching only $\eta\!\approx\!0.60$ at a $50\%$ budget) and redirects the fix to the feature axis (\S\ref{sec:shape}). Low-rank/SVD on KV~\citep{gear2024,sun2025shadowkv} targets compression/quantization of one context, not cross-chunk binding; Semantic Cache Distillation~\citep{scd2026} learns a low-rank aligner for cross-\emph{model} drift. Dynamic sparse attention (StreamingLLM~\citep{xiao2024streamingllm}, H2O~\citep{zhang2023h2o}, Quest~\citep{tang2024quest}) selects tokens \emph{within} a context to cut decode bandwidth, not to restore antecedent conditioning. The low-rank-patch \emph{mechanism} and the ``correct at depth'' observation are prior; what is new here is the diagnosis that redirects the fix, the unification across MLA/GQA/MHA, and the position/conditioning separation that makes reuse recompute-free and eviction reversible.

\section{Conclusion}

A multimodal agent's context outgrows its window through patterns a prefix cache cannot serve: sliding windows, reorderings, look-backs. The recompute is avoidable: the only thing reuse loses is cross-chunk conditioning, diffuse in tokens but low-rank in features and deep, so a position-free canonical plus a rank-$m$ patch reconstructs a chunk's KV at any position with one operator across MLA/GQA/MHA. This makes reorder free over an orbit, window slides free for survivors, and eviction reversible: recall costs a single patch on the fixed past, never a re-encode, reconstructed to within bf16 rounding in a live engine.

More broadly, once a chunk's binding is a small additive object rather than a baked-in recompute, the KV cache stops being a position-indexed array and becomes a structure an orchestrator edits cheaply: reversible eviction, context-clean forks, content-addressed reuse, and---because reorder is free over an orbit---reuse-aware \emph{placement}, where a window's contents are a set and chunk order becomes a scheduling variable rather than a consequence of arrival (\S\ref{app:ctx}). The recall cascade, this placement problem, and the workloads where reuse amortizes are left to future work. Context beyond the window need not be context recomputed, nor the model retrained to serve it.

\ifcolmsubmission\else
\section*{Acknowledgments}
The authors gratefully acknowledge the scientific support and HPC resources provided by the Erlangen National High Performance Computing Center (NHR@FAU) of the Friedrich-Alexander-Universit\"at Erlangen-N\"urnberg (FAU). The hardware is funded by the German Research Foundation (DFG).
\fi

\bibliography{kamera}
\bibliographystyle{colm2026_conference}

\appendix
\section{Forming and applying the patch}
\label{app:impl}

The operator splits into a compile step (once per chunk, amortized) and a serve step (once per reuse, forward-free). \textsc{Compile} runs a single conditioned forward over $[\,\text{prefix}\cdot A\cdot B\,]$, reads $B$'s conditioned KV, subtracts the stored relocated canonical to obtain the deficit $\Dl$, and keeps its top-$m$ SVD factors ($\approx\!2\%$ of the page):
{\small
\begin{verbatim}
def form_patch(prefix, A, B, delta, m, layer):     # COMPILE: once per chunk, amortized
  kv_cond  = forward(concat(prefix, A, B)).kv[B, layer]   # KV(B|A): one conditioned forward
  kv_solo  = rotate_rope(stored.content[B, layer], delta) # R(delta).KV(B|emptyset), cached
  Delta    = kv_cond - kv_solo                            # cross-chunk conditioning deficit
  U, S, Vt = svd(Delta)                                          # keep top-m factors only
  return U[:, :m] * S[:m], Vt[:m]                         # stored patch {U_m, V_m} (~2% of page)
\end{verbatim}}
\textsc{Serve} applies Eq.~\ref{eq:operator} with zero forwards: a per-layer RoPE rotation of the stored keys to the matched position, then a rank-$m$ GEMM into the paged KV cache. It is bandwidth-bound, rank-invariant in latency, and needs no engine surgery beyond a cache hook; only the stored content KV and the small factors are read from HBM, and the vision encoder and the chunk's prefill are skipped entirely:
{\small
\begin{verbatim}
def apply_reuse(stored, delta, U, V, layer):          # SERVE: per reused chunk, per layer
  K, Vv = stored.content_K[layer], stored.content_V[layer]        # KV(B|emptyset), bf16
  rotate_rope_inplace(stored.rope_band[layer], delta)        # R(delta): exact, V untouched
  K  = K  + U.K[layer] @ V.K[layer].T                        # rank-m conditioning patch (K)
  Vv = Vv + U.V[layer] @ V.V[layer].T                        # both channels carry binding (V)
  return assemble(K, Vv, stored.rope_band[layer])                   # -> FlashAttention-3
\end{verbatim}}

\section{A menu of cross-chunk reuse operating points and its boundary}
\label{app:menu}

The recompute-free path is one point on a spectrum graded by how tightly the new request relates to what is cached: free and exact when the chunk is \emph{leading} (deficit $0$, a radix hit), a single reused \emph{orbit}-patch when the request only reorders the cached set, an amortized millisecond patch when the antecedent recurs, and the one-time forming cost for a never-seen antecedent (Table~\ref{tab:menu}). The floor is a dominating operating point: the worst case is to re-prefill, so reuse through the stored canonical is never worse than recompute in fidelity (it reconstructs re-prefill KV to within bf16 rounding) nor, once formed, in cost. A prefix/embedding cache expresses only the leading, identical-order lane; the canonical store opens the rest.

\begin{table}[h]
\centering\footnotesize\setlength{\tabcolsep}{3.5pt}
\begin{tabular}{l l l l}
\toprule
operating point & mechanism & cost & where it holds \\
\midrule
leading-segment reuse & truncate $+$ relocate & free (deficit $=0$) & all (causal) \\
exact per-context & rank-$m$ SVD$(\Dl)$ & one $B$-forward & all (ceiling) \\
reorder orbit-patch & one patch / orbit & one patch / chunk & all families \\
deep-half patch & deepest $\sim\!n_L/2$ only & half the bytes & all ($\sim\!95\%$) \\
reversible eviction & relocate / patch recall & free / rank-$32$ patch & GQA/deepst./MLA \\
partial forward & shallow reuse, deep recompute & $\ell^\star/n_L$ of prefill & depth-bounded \\
\bottomrule
\end{tabular}
\caption{Operating points for cross-chunk KV reuse, graded by the request--cache relationship. \emph{Ruled out at the boundary} (cheap-estimation negatives that bound the menu's shape): the attention-sink prosthesis, a per-antecedent linear operator, key-similarity selection, A-side streaming, and a shallow-seed predictor of the deep patch, all failing because the deficit's coefficients are item-specific even though its directions are universal.}
\label{tab:menu}
\end{table}

\begin{figure}[h]\centering\includegraphics[width=0.86\linewidth]{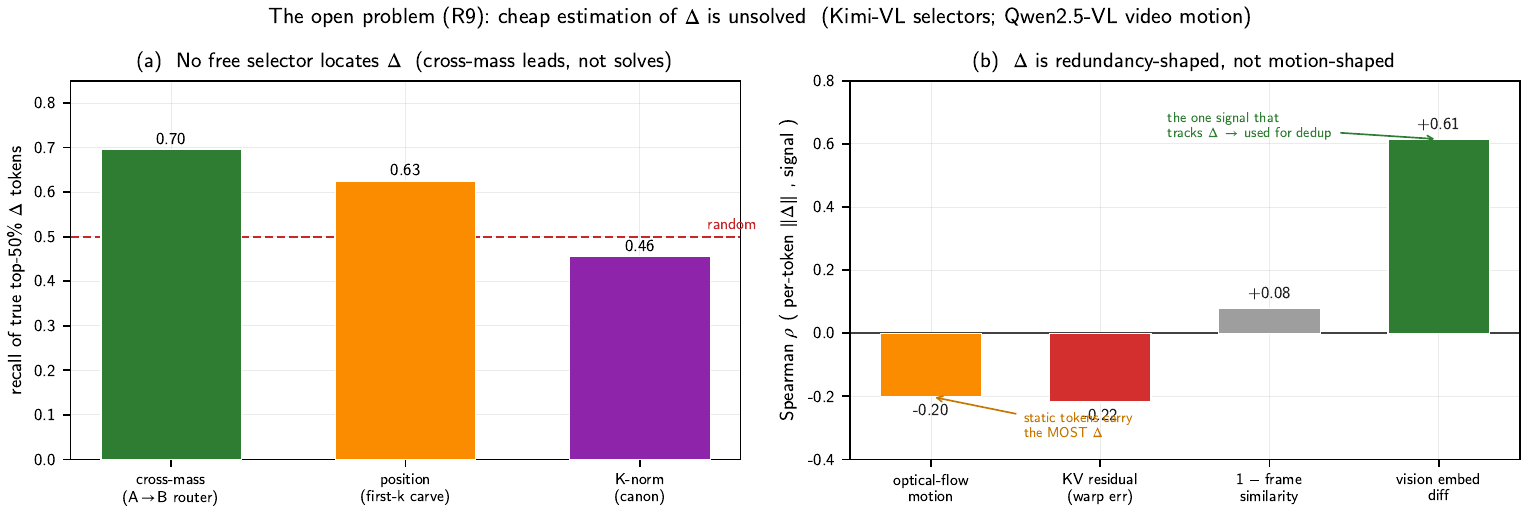}\caption{Why cheap estimation of the deficit is still open. (a) no cheap selector locates it---the best signal (cross-attention mass) reaches $0.70$ recall but does not solve it; (b) the deficit is redundancy-shaped: anti-correlated with motion and uncorrelated with frame similarity, so only an actual vision-embedding difference tracks it.}\label{fig:cheap}\end{figure}

The lower boundary of the menu is the open problem of estimating $\Dl$ \emph{cheaply}, without the conditioned forward (Fig.~\ref{fig:cheap}): no free selector locates it, and it is redundancy-shaped (anti-correlated with motion, uncorrelated with frame similarity), so every cheap content-change signal mispredicts it. The directions of $\Dl$ are model-intrinsic and free to share, but the per-token coefficients require observing $B$ attend $A$ under the conditioned forward.

\paragraph{Eviction-probe details.} The recall sweep (Table~\ref{tab:evict}) serves a bounded window of $k$ chunks over a growing history at turnover fractions $\tau\in\{0,0.2,\dots,1.0\}$; ``stale'' replays the patch frozen at eviction, ``fresh'' re-forms a rank-$32$ patch on the chunk's earlier (now fixed) context. The survivor probe evicts the leading chunk and measures the surviving interior chunk under keep-as-is ($\rd$ only) versus a rank-$r$ removal patch; $d_{\text{remove}}$ relative norm is reported by depth, with the $90\%$-energy rank establishing that a rank-$64$ removal patch is sufficient. Both probes run on cached Video-MME segments with $n{=}25$--$32$ source clips per model across GQA (Qwen2.5-VL, $n_L{=}28$), deepstack-GQA (Qwen3-VL, $n_L{=}36$), and MLA (Kimi-VL, $n_L{=}27$).

\section{Supporting evidence for the feature patch}
\label{app:obs}

The body foregrounds the three window operations. Here we collect the evidence behind its claims, in the order they build the argument: blind reuse breaks multi-hop accuracy while the patch restores it (\S\ref{app:obs:acc}); the deficit is intrinsically thin (\S\ref{app:obs:thin}), which is why the token, layer, and head selection of prior work misses (\S\ref{app:obs:miss}); the deficit and its repair are architecture-universal (\S\ref{app:obs:arch}); the effect lives in redundant streams and vanishes for text (\S\ref{app:obs:scope}); and the operator deploys at the bf16 reconstruction floor with a memory win on a live engine (\S\ref{app:obs:deploy}).

\subsection{Reuse breaks multi-hop accuracy; the patch restores it}
\label{app:obs:acc}
The distinction from single-hop-readout and single-image prior work is a ground-truth accuracy gap that opens \emph{only} where the answer needs cross-chunk binding. On MM-NIAH retrieval-image (a single image needle that must bind to the query, not readout-decomposable) and on the multi-hop reasoning-image split, blind KV reuse halves accuracy and a rank-16 conditioning patch restores the re-prefill ceiling --- on \emph{both} Qwen2.5-VL (GQA) and Kimi-VL (MLA), the two clean KV families (Table~\ref{tab:mmniah}). Next-token KL to re-prefill collapses correspondingly (rank-64: $0.61\!\to\!0.009$ on Qwen retrieval, $0.087\!\to\!0.006$ on Kimi reasoning). On two-page multi-hop document QA the feature patch reaches the re-prefill ceiling on both MLA and GQA at a few MB, while the token-axis selectors the literature uses fall well short at matched budget (Table~\ref{tab:headline}).

\begin{table}[h]
\centering\small
\begin{tabular}{lcccc}
\toprule
 & \multicolumn{2}{c}{Qwen2.5-VL (GQA)} & \multicolumn{2}{c}{Kimi-VL (MLA)} \\
\cmidrule(lr){2-3}\cmidrule(lr){4-5}
accuracy $\uparrow$ & reason. & retr. & reason. & retr. \\
\midrule
re-prefill (ceiling)   & 0.59 & 0.74 & 0.80 & 0.55 \\
blind reuse            & 0.41 & 0.38 & 0.57 & 0.35 \\
$+$ rank-16 patch      & 0.64 & \textbf{0.72} & 0.79 & 0.52 \\
$+$ rank-64 patch      & 0.59 & 0.70 & \textbf{0.82} & \textbf{0.55} \\
\midrule
chance / $n$           & .50/56 & .25/61 & .50/56 & .25/29 \\
\bottomrule
\end{tabular}
\caption{Ground-truth accuracy across two KV families (GQA, MLA) on two MM-NIAH image tasks that both demand cross-chunk binding: \emph{retrieval-image} (one needle binding to the query) and the harder multi-hop \emph{reasoning-image} split. Blind reuse drops accuracy toward chance on every model$\times$task; the rank-$m$ conditioning patch restores it to the re-prefill ceiling and beats the token baselines (CacheBlend/sink, e.g.\ Kimi reasoning $0.55$--$0.71$ vs.\ patch $0.82$). Next-token KL to re-prefill collapses in step (rank-64: Qwen retrieval $0.61\!\to\!0.009$, Kimi reasoning $0.087\!\to\!0.006$). Kimi retrieval $n{=}29$ (the processor's image-run detection dropped items); the multi-hop reasoning cells are $n{=}56$.}
\label{tab:mmniah}
\end{table}

\begin{table}[h]
\centering\small
\begin{tabular}{llcc}
\toprule
recovery axis & scheme & MLA multi-hop & GQA multi-hop\\
\midrule
--- & re-prefill (ceiling) & 0.41 & 0.28\\
--- & blind reuse (floor) & 0.28 & 0.15\\
\textbf{feature (ours)} & \texttt{lm\_16} ($\approx\!6\%$ KV) & 0.39 & 0.24\\
\textbf{feature (ours)} & \texttt{lm\_64} ($\approx\!25\%$ KV) & \textbf{0.41} & \textbf{0.28}\\
token selector & cross-mass $0.25$ & 0.35 & ---\\
token subset & leverage-Ny\"strom $64$ & 0.26 & ---\\
\bottomrule
\end{tabular}
\caption{The contribution-defining comparison on two-page multi-hop document QA: the feature-axis patch reaches the re-prefill ceiling at a small fraction of the segment's KV, while the token-axis methods the literature uses do not, at matched budget. Byte fractions ($\approx\!6\%/25\%$ at rank-16/64) are layout-invariant: the same on the per-head $K/V$ page and on the MLA latent ($c_{KV}{+}k_{pe}$) page, since patch and page scale together.}
\label{tab:headline}
\end{table}

\subsection{The deficit is low-rank, so the patch is thin}
\label{app:obs:thin}
\begin{figure}[h]
\centering
\includegraphics[width=0.6\linewidth]{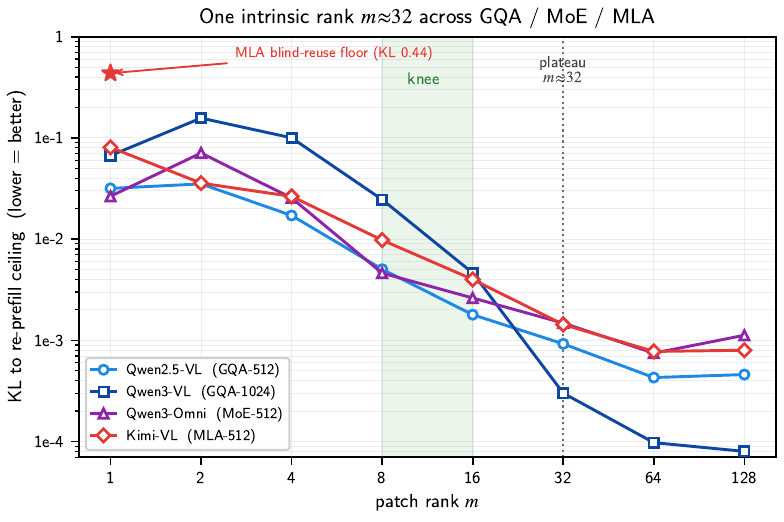}
\caption{One intrinsic rank (about $32$) governs the correction across GQA, MoE, and MLA and across $512$ vs.\ $1024$ hidden width---the saturating rank is a property of the model, not of its width.}
\label{fig:rank}
\end{figure}

Sweeping $m$ on a multi-image workload, the conditioning-KL knees at $m\!\approx\!8$--$16$ and plateaus by $m\!\approx\!32$ on \emph{every} structure (GQA-512, GQA-1024, MoE, MLA; Fig.~\ref{fig:rank}). Rank-32 closes $94\%$ of the KL gap to the rank-64 floor; a held-out split selects the same plateau (bootstrap $95\%$ CI $[32,128]$). The \emph{saturating} rank is absolute, not a width fraction: the $1024$-wide model plateaus at the same $m$ as the $512$-wide one (the curves coincide at the plateau; below it the wider model trails, Fig.~\ref{fig:rank}). The directions are moreover shared across items (a fixed pooled basis recovers a held-out deficit as well as the item's own SVD, \S\ref{sec:shape}), so the patch is not only thin but reusable: one basis serves many chunks, and only the per-token coefficients are item-specific.

\subsection{Why token, layer, and head selection miss}
\label{app:obs:miss}
\begin{figure}[h]
\centering
\includegraphics[width=\linewidth]{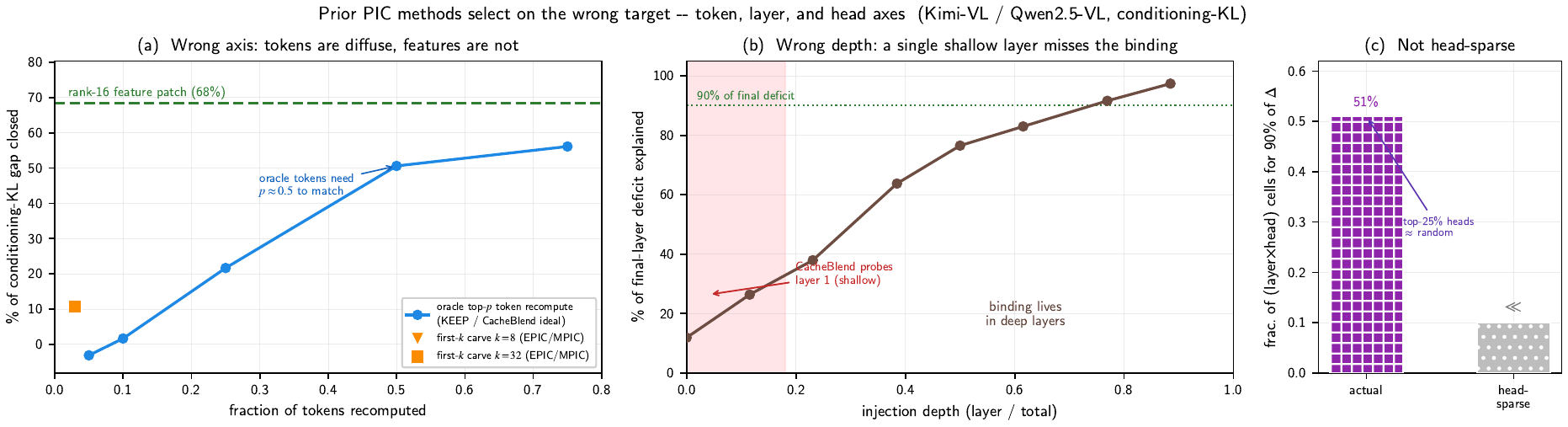}
\caption{Prior position-independent caches select on the wrong target. (a) \emph{wrong axis}: an oracle that recomputes tokens needs about half of them and a first-$k$ carve recovers almost nothing, while a rank-16 \emph{feature} patch closes most of the gap; (b) \emph{wrong depth}: a single shallow layer explains little of the final deficit; (c) \emph{wrong grouping}: the deficit is not concentrated in a few attention heads.}
\label{fig:wrong}
\end{figure}

Because the deficit is low-rank in features but diffuse in tokens and concentrated deep, the token/layer/head-selection premise of position-independent caching misses on all three axes (Fig.~\ref{fig:wrong}). \emph{Wrong axis}: oracle top-$p$ token recompute needs $p\!\approx\!0.5$ and a first-$k$ carve recovers $\approx\!0$, while a rank-16 feature patch closes $68\%$. \emph{Wrong depth}: a single shallow layer explains little of the final deficit. \emph{Wrong grouping}: $\Dl$ is not head-sparse ($90\%$ of its energy needs $\approx\!51\%$ of (layer$\times$head) cells). At a matched KV-byte budget the feature patch closes $82/90\%$ of the loss at rank-16/64 versus an oracle query-aware (Quest-style) selector's $55/79\%$, first-$k$'s $31/41\%$, and low-rank-$K$'s $\approx\!0$ (Table~\ref{tab:matched}). The same failure shows up downstream on a real multi-hop video task (Table~\ref{tab:milebenchT}): the rank-64 patch recovers $97\%$ of the answer flips recompute-free, while VLCache and CacheBlend stay near the blind floor at KL $>\!1.1$, VLCache no better than a uniform attention sink; only a shallow partial re-prefill keeps up, at the cost of an in-context forward.

\begin{figure}[h]
\centering
\includegraphics[width=\linewidth]{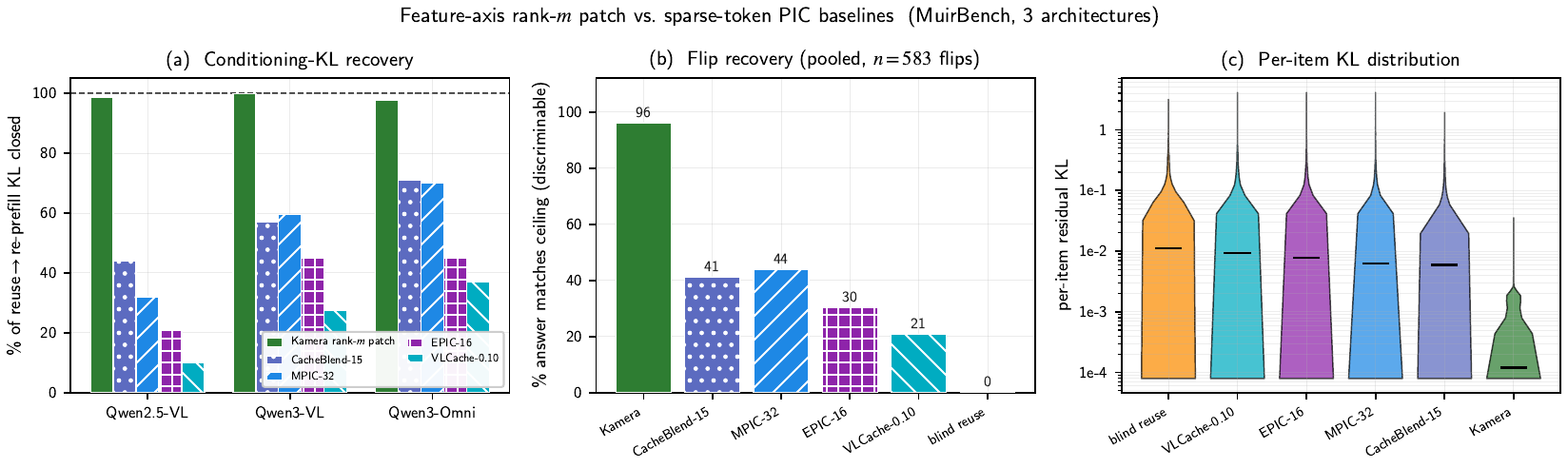}
\caption{Feature-axis patch vs.\ sparse-token PIC baselines (MuirBench~\citep{wang2024muirbench}, three architectures), all given the same relocated KV. \textbf{(a)} how much of the conditioning gap each closes---the patch nearly all of it ($98$--$100\%$), the token baselines a fraction ($10$--$71\%$); \textbf{(b)} how often each restores the re-prefill answer on the items blind reuse flips---the patch almost always ($96\%$), the token baselines rarely ($21$--$44\%$), since closing the gap is necessary but not sufficient; \textbf{(c)} per-item residual error: the patch hugs zero while every token baseline sits near blind reuse.}
\label{fig:baselines}
\end{figure}

\begin{table}[h]
\centering\small
\begin{tabular}{lccccc}
\toprule
MileBench temporal (Qwen2.5-VL), flip-recover $\uparrow$ & AS & AP & OS & SC & mean KL $\downarrow$ \\
\midrule
re-prefill (ceiling)                       & 1.00 & 1.00 & 1.00 & 1.00 & 0.00 \\
$+$ rank-64 conditioning patch (ours)      & \textbf{1.00} & \textbf{1.00} & \textbf{0.94} & \textbf{0.93} & \textbf{0.013} \\
VLCache-0.10 (uniform $10\%$ KV keep)       & 0.10 & 0.04 & 0.06 & 0.00 & 1.23 \\
CacheBlend-15 (max-deviation token)        & 0.25 & 0.09 & 0.11 & 0.36 & 1.13 \\
attention-sink prosthesis ($k{=}32$)       & 0.10 & 0.04 & 0.11 & 0.21 & 1.22 \\
shallow re-prefill (bottom $3$ layers)     & 0.90 & 0.96 & 0.89 & 0.93 & 0.027 \\
\bottomrule
\end{tabular}
\caption{Answer-flip recovery on four MileBench temporal benchmarks (AS\,$=$\,ActionSequence, AP\,$=$\,ActionPrediction, OS\,$=$\,ObjectShuffle, SC\,$=$\,StateChange; pooled flip subset $n{=}75$). On items where blind reuse flips the answer, the rank-$64$ patch recovers the re-prefill decision recompute-free; token-recompute baselines fail (VLCache no better than a uniform sink); only a shallow partial re-prefill keeps up, at the cost of an in-context forward. VLCache uses a uniform per-layer keep budget here, which understates its layer-adaptive schedule (a conservative test).}
\label{tab:milebenchT}
\end{table}

\begin{table}[h]
\centering\small
\begin{tabular}{lcc}
\toprule
recovery at matched KV-byte budget & rank-16 ($\equiv\!31$ tok) & rank-64 ($\equiv\!124$ tok) \\
\midrule
feature patch (ours, low-rank $\Dl$)            & \textbf{0.82} & \textbf{0.90} \\
oracle query-aware recompute (Quest-style)      & 0.55          & 0.79 \\
first-$k$ recompute (EPIC/MPIC-style)            & 0.31          & 0.41 \\
low-rank-$K$ reconstruction (ShadowKV-style)     & $\le 0$       & $\le 0$ \\
\bottomrule
\end{tabular}
\caption{Fraction of the multi-hop conditioning loss closed at a \emph{matched} KV-byte budget (Qwen2.5-VL, $n{=}46$). The feature patch exceeds every token-axis recovery: decisively over first-$k$ (EPIC/MPIC) and low-rank-$K$ (ShadowKV), and with a paired-significant margin over an oracle query-aware selector (rank-64 paired $\Delta\eta{=}0.24$, $95\%$ CI $[0.07,0.41]$), establishing that the token/page axis, not just a specific selector, is the wrong one. Recomputed tokens re-attend the full context (strongest CacheBlend form).}
\label{tab:matched}
\end{table}

\begin{figure}[h]
\centering
\includegraphics[width=0.7\linewidth]{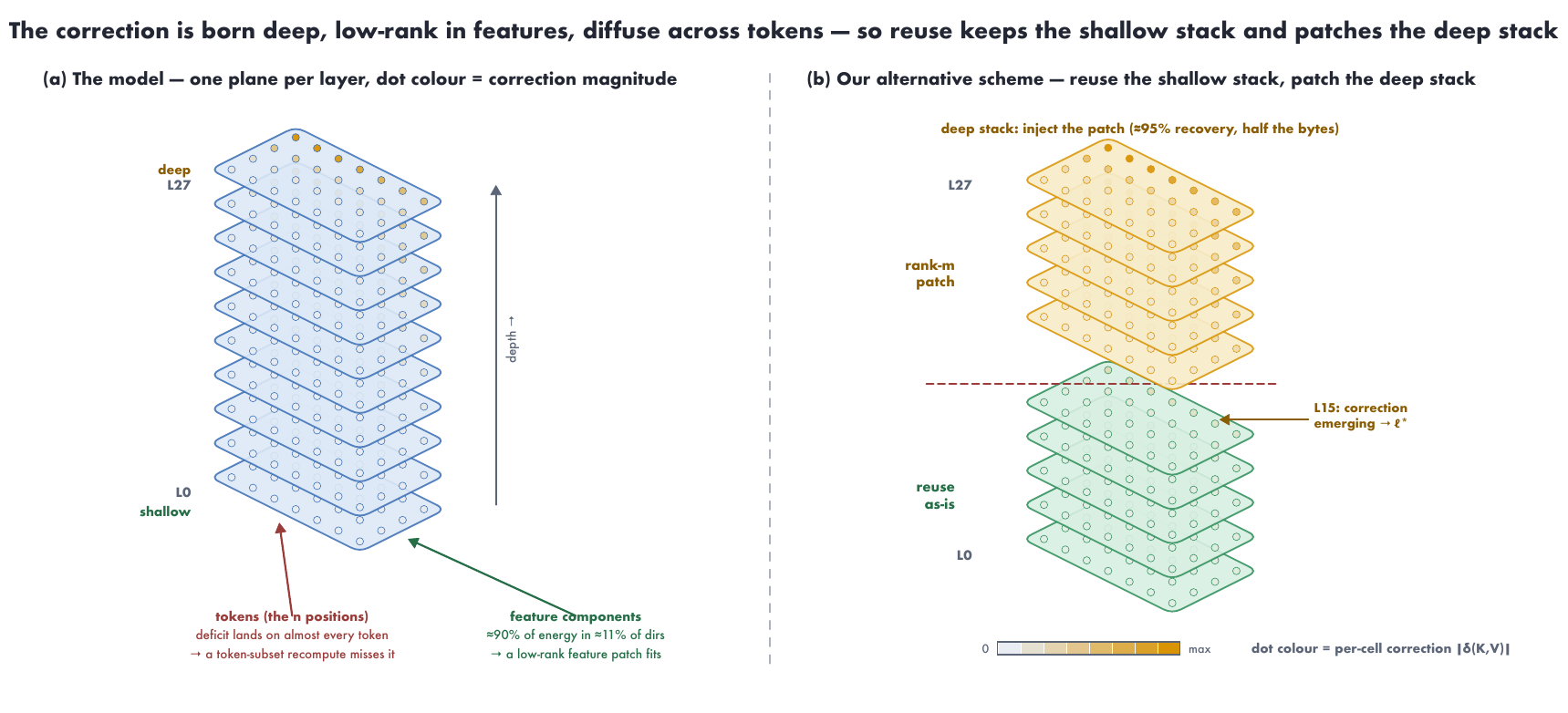}
\caption{The deficit is shallow-free and deep. Each plane is one layer's feature-by-token map: shallow layers reuse verbatim, the deficit grows from the middle layers downward, and it spreads across tokens rather than concentrating in a few---so neither a token subset nor a single shallow layer captures it. A low-rank patch on the deep layers restores it; equivalently, recomputing only the deep layers in context keeps up.}
\label{fig:iso}
\end{figure}

The depth structure underneath these results is shown in Fig.~\ref{fig:iso}: shallow layers carry the chunk's context-free representation and reuse verbatim, while the conditioning deficit emerges in the middle layers and concentrates in the deep ones, spreading across tokens rather than into a few columns. This is why a single shallow layer or a token subset cannot localize it, why the correction is a low-rank patch on the deep layers, and why the one partial-recompute lever that keeps up is a \emph{shallow re-prefill}, reusing the shallow layers and recomputing only the deep, entangled ones in context. The depth structure also makes the patch itself layer-sparse: storing only the deepest $\sim\!n_L/2$ layers' factors halves the patch bytes at $\sim\!95\%$ of full fidelity (Table~\ref{tab:menu}), a cheaper alternative whose depth budget is \emph{model-dependent}, shallower for dense VLMs and deeper for the deepstack backbone whose visual re-injection pushes binding down (\S\ref{sec:window}). We credit the ``correct at depth'' observation to prior work (\S\ref{sec:related}); what we add is this layer-sparse storage lever and its per-model budget, not the depth finding itself.

\subsection{The deficit and its repair are architecture-universal}
\label{app:obs:arch}
\begin{figure}[h]
\centering
\includegraphics[width=\linewidth]{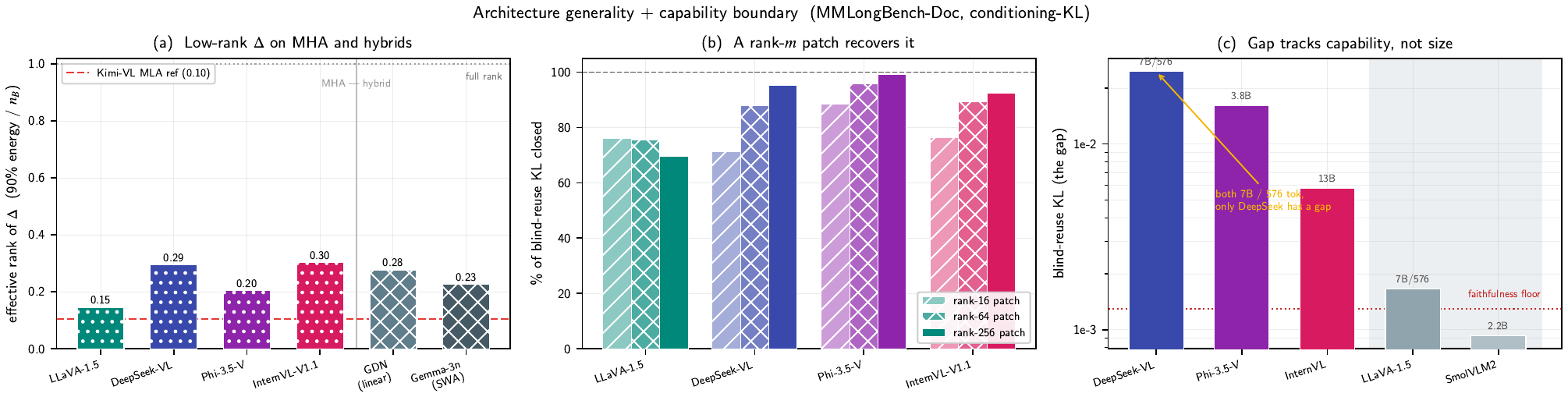}
\caption{Low-rank $\Dl$ and monotone rank-$m$ recovery across softmax-attention VLMs (pure-MHA and KV-sharing variants); the gap tracks cross-chunk-binding \emph{capability}, not size (LLaVA-1.5 and DeepSeek-VL are both 7B/576-tok, only DeepSeek shows a gap).}
\label{fig:archgen}
\end{figure}

The cross-chunk deficit (the cross-chunk-binding loss, since single-hop reads are exactly recovered) holds across six backbones, isolated directly with the 4D mask of \S\ref{sec:bg}: block $B\!\not\to\!A$ at $B$'s native positions, so the residual is conditioning with zero position contribution by construction (Table~\ref{tab:condarch}). Over each model's full valid probe set ($n{=}46$--$94$), the position-matched control sits $10$--$320\times$ below the conditioning loss, the deficit is low-rank ($e_{90}/n_B\!\le\!0.30$), and a rank-64 patch closes $85$--$96\%$ of the per-item deficit ($92$--$99\%$ at rank-256), so the result is not an artifact of one model or of small $n$. The recovery is monotone in rank on three pure-MHA VLMs as well (Fig.~\ref{fig:archgen}); the gap is \emph{absent} in the weakest models (near-zero for SmolVLM2~\citep{marafioti2025smolvlm}, LLaVA-1.5~\citep{liu2024llava15}) and present across capable backbones, a single-axis observation (the deficit requires a model that actually binds across chunks), not a clean function of size. Two architectural axes are covered. Along the attention / KV-sharing axis (MLA's latent vs.\ GQA's grouped heads vs.\ MHA's full per-head keys) one operator applies, as above. Along the FFN-sparsity axis (dense vs.\ MoE) the deficit is unchanged: it is an \emph{attention} object, so the MoE backbone (Qwen3-Omni~\citep{xu2025qwen3omni}) recovers like a dense model (rank-32 closes $\ge\!91\%$, Fig.~\ref{fig:rank}), since routing lives in the FFN while binding lives in attention. All families here are KV-sharing variants of \emph{softmax} attention; a linear-attention or SSM layer carries no KV to patch (its analogue is a state-delta) and is outside this operator's scope.

\begin{table}[h]
\centering\small
\begin{tabular}{lccccc}
\toprule
model (family) & $n$ & ctrl-KL & loss-KL & $e_{90}/n_B$ & gap@64 \\
\midrule
Qwen2.5-VL (GQA)        & 48 & 0.0025 & 0.025 & 0.20 & 0.85 \\
Qwen3-VL (GQA-interl.)  & 94 & 0.0001 & 0.001 & 0.26 & 0.92 \\
InternVL3-8B~\citep{zhu2025internvl3} (GQA)      & 46 & 0.0029 & 0.046 & 0.18 & 0.89 \\
DeepSeek-VL (MHA)       & 46 & 0.0009 & 0.025 & 0.29 & 0.87 \\
InternVL-V1.1~\citep{chen2024internvl} (MHA)     & 46 & 0.0005 & 0.006 & 0.30 & 0.89 \\
Phi-3.5-V~\citep{abdin2024phi3} (MHA)         & 46 & 0.0001 & 0.016 & 0.20 & 0.96 \\
\bottomrule
\end{tabular}
\caption{The cross-chunk conditioning deficit holds across six backbones (4D-mask isolation, $B$ at native positions; this isolation set and the repair-frontier Table~\ref{tab:frontierarch} each report six, eleven models in union across all experiments). The position-matched control (ctrl-KL) sits $10$--$320\times$ below the conditioning loss (loss-KL), $\Dl$ is low-rank ($e_{90}/n_B\!\le\!0.30$ on $V$, lower still on GQA's $K$), and a rank-64 patch closes $85$--$96\%$ of the per-item deficit ($92$--$99\%$ at rank-256). Position is handled separately and exactly by $\rd$ (Fig.~\ref{fig:deploy}b); the MoE backbone (Qwen3-Omni) shows the same low-rank recovery on the reuse axis (Figs.~\ref{fig:rank},~\ref{fig:baselines}). \texttt{gap@64} is the median over items with a measurable deficit of the fraction of loss-KL a rank-64 patch closes.}
\label{tab:condarch}
\end{table}

The repair frontier is architecture-universal too (Table~\ref{tab:frontierarch}): across MLA, GQA, deepstack-GQA, MoE, and pure MHA the feature patch reaches near-ceiling fidelity at \emph{zero} LLM-prefill recompute, while layer re-prefill must re-run $\sim\!86$--$89\%$ of layers to match it and token recompute saturates well below the patch at a $50\%$ budget.

\begin{table}[h]
\centering\small
\setlength{\tabcolsep}{4.5pt}
\begin{tabular}{l l c c c}
\toprule
model & family ($n_L$) & patch $\eta$@$0$ & re-prefill to match & token $\eta$@$0.5$ \\
\midrule
Kimi-VL-A3B    & MLA ($27$)           & $0.91$ & $0.89$ & $0.32$ \\
Qwen2.5-VL-7B  & GQA ($28$)           & $0.96$ & $0.86$ & $0.60$ \\
Qwen3-VL-8B    & deepstack-GQA ($36$) & $0.88$ & $0.89$ & $0.63$ \\
Qwen3-Omni-30B & MoE ($48$)           & $0.90$ & $0.88$ & $0.86$ \\
DeepSeek-VL-7B & MHA ($30$)           & $0.96$ & $\sim\!1.0$ & $0.70$ \\
Phi-3.5-V      & MHA ($32$)           & $0.90$ & $0.88$ & $0.68$ \\
\bottomrule
\end{tabular}
\caption{The repair frontier is architecture-universal ($n{=}40$/model, $\eta$ at rank-$64$). The feature patch reaches near-ceiling fidelity at \emph{zero} LLM-prefill recompute on every family; layer re-prefill must re-run $\sim\!86$--$89\%$ of layers (essentially all on DeepSeek-VL) to match it; token recompute saturates well below the patch even at a $50\%$ budget. ``Re-prefill cost to match'' is the re-run fraction at which a partial forward first reaches the patch's $\eta$.}
\label{tab:frontierarch}
\end{table}

\subsection{Where the effect lives: modality scope}
\label{app:obs:scope}
\begin{figure}[h]\centering\includegraphics[width=0.92\linewidth]{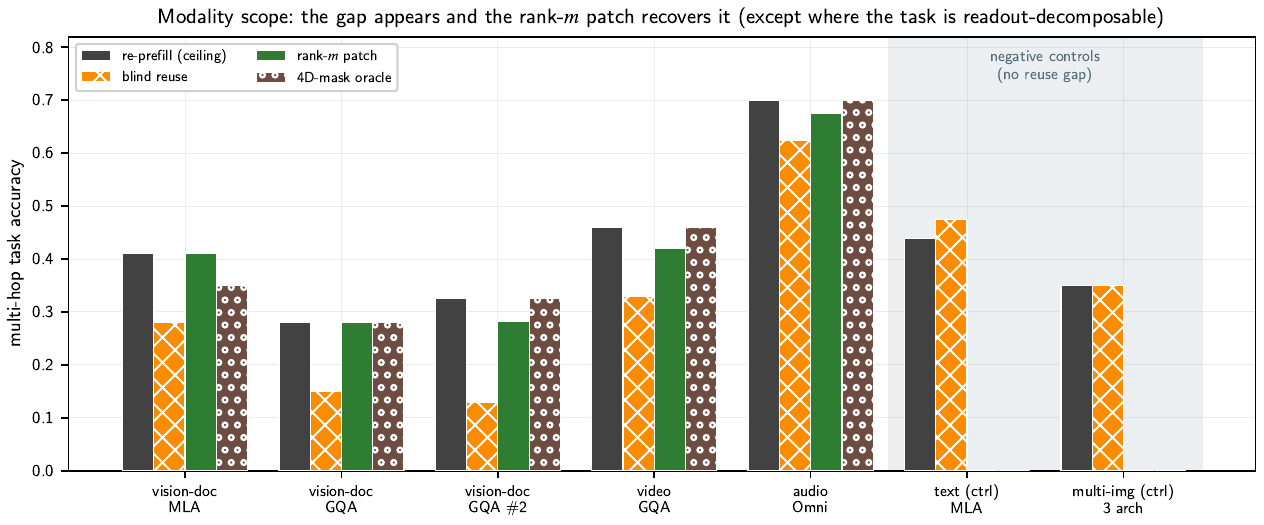}\caption{Modality scope: vision and video show the gap and recover; audio (Qwen2.5-Omni, $n{=}40$, exploratory) a smaller partial gap; dense text and readout-decomposable multi-image MC are negative controls.}\label{fig:modality}\end{figure}

The deficit is a property of \emph{redundant} token streams whose meaning lives in cross-chunk binding. Vision and video show the gap and recover; audio shows a smaller, partly-recovered gap (Qwen2.5-Omni, $n{=}40$, within-noise and treated as exploratory); dense text (MuSiQue 2-hop) shows none, and readout-decomposable multi-image MC is a negative control (Fig.~\ref{fig:modality}). Document \emph{images} of text lose binding while the same facts as text tokens do not, so the effect is concentrated in the multimodal context a windowed agent accumulates, and absent where prefix caching of text already suffices.

\subsection{Memory cost and bf16-faithful live deployment}
\label{app:obs:deploy}
\begin{figure}[h]
\centering
\includegraphics[width=\linewidth]{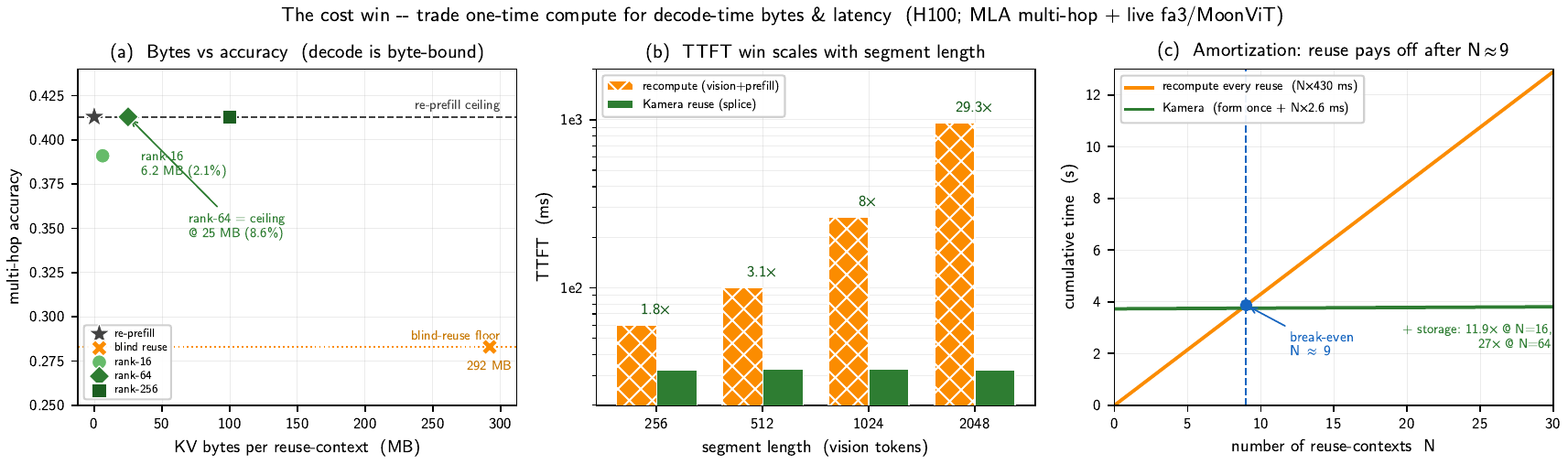}
\caption{The cost is on memory. (a) decode sits far on the bandwidth-bound side of the H100 roofline, so KV bytes are the binding resource; (b) replacing the per-reuse prefill with a forward-free patch-apply improves time-to-first-token by up to $29\times$ as the reused segment grows ($256\!\to\!2048$ tokens; larger still against full recompute, which also re-runs the vision encoder); (c) the one-time forming forward is repaid after about $9$ reuses, against re-prefilling each time.}
\label{fig:cost}
\end{figure}

\begin{figure}[h]
\centering
\includegraphics[width=\linewidth]{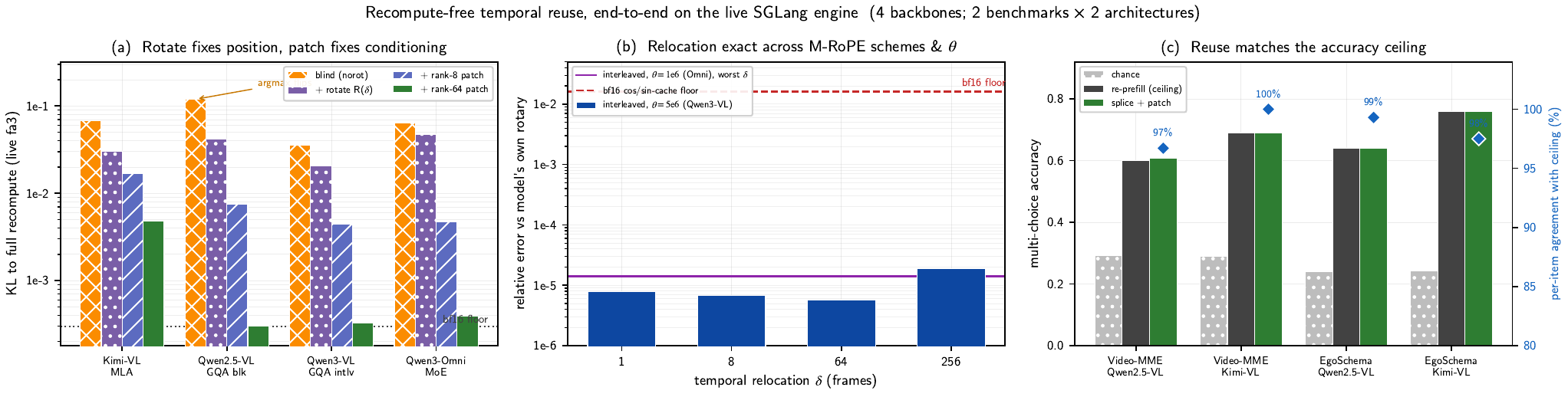}
\caption{Recompute-free temporal reuse in the live SGLang kernel. (a) the operator's error over four backbones: re-rotation fixes position and the patch fixes conditioning, bringing the reconstructed KV to within one bf16 unit of recompute (residual next-token KL about $10^{-3}$, far below blind reuse); (b) relocation stays exact across the different M-RoPE layouts and rotary bases; (c) downstream multiple-choice accuracy equals the re-prefill ceiling, with per-item agreement shown on the right.}
\label{fig:deploy}
\end{figure}

Decode is bandwidth-bound (Fig.~\ref{fig:cost}a), so the binding resource is KV bytes; a rank-64 patch matches full multi-hop accuracy at $\approx\!25\%$ of the segment's KV bytes (layout-invariant; rank-16 $\approx\!6\%$) and the forming forward amortizes after $\approx\!9$ reuses against a prefill-per-reuse baseline (Fig.~\ref{fig:cost}c), with the LM prefill replaced by patch-apply improving TTFT $1.8\times\!\to\!29\times$ as segments grow $256\!\to\!2048$ tokens (Fig.~\ref{fig:cost}b). On the live SGLang engine the operator's KV reconstruction descends to the bf16 floor across four backbones (residual logit KL $\approx\!10^{-3}$), relocation is exact across M-RoPE schemes, and downstream MC accuracy equals the re-prefill ceiling (Fig.~\ref{fig:deploy}).

\section{The reuse safety envelope: when a cached patch survives context drift}
\label{app:safety}

A windowed agent's context drifts: the antecedent in front of a cached chunk is reordered, partly replaced, or grown with new material as the window slides. A stored patch is conditioned on a specific antecedent, so the operating question is how far that antecedent can drift before the cached patch must be rebuilt. We bound this on three controlled perturbations of the predecessor set, holding $B$ fixed.

\paragraph{Divergent antecedents.} Perturbing $A\!\to\!A'$ at matched positions (reorder; drop-and-duplicate a frame; replace one or all predecessors with frames from a different clip), the stored patch transfers gracefully on Qwen2.5-VL: a dropped or duplicated frame gives $\eta_{\text{transfer}}{=}0.92$, indistinguishable from recomputing the patch ($\eta_{\text{exact}}{=}0.92$); reorder and single-replace hold at $0.76$--$0.77$; only when \emph{all} predecessors become a different clip (cosine divergence $0.43$) does the stale patch turn harmful ($-3.5$) while the exact patch still recovers ($0.95$). The decay is graceful and tracks divergence, in contrast to a prefix cache's step-to-zero at the first differing token. As a \emph{binary} reuse-vs-rebuild gate the cheap divergence signal is too weak (false-reuse $24$--$47\%$ at any useful coverage), but the safe/unsafe divergence medians separate ($0.016$ vs.\ $0.12$ on GQA, $0.008$ vs.\ $0.024$ on MLA), so divergence is a real but soft prioritization hint.

\paragraph{Excess context (the superset boundary).} When the served context carries \emph{extra} irrelevant material, the rule is to rebuild rather than reuse. Position-matched, the stale $\{X,Y\}$ patch decays steadily as $d$ distractor frames accumulate ($\eta_{xy}{=}0.93\!\to\!0.71\!\to\!0.39\!\to\!0.08$ over $d{=}0$--$3$ on Qwen2.5-VL) while a patch rebuilt with the distractors present stays flat ($\approx\!0.92$): the patch operator is distractor-agnostic, only its \emph{staleness} hurts. On the conditioning-bound flip subset the stale patch loses the decision (flip-recover $1.0\!\to\!0.17$) while the rebuilt patch holds ($\ge\!0.83$). Under a true superset (count growth plus relocation, MLA) the stale patch turns actively harmful by $d{\ge}2$, and the tolerance shrinks with depth: on the deepstack backbone the stale patch already goes negative at $d{=}2$ even position-matched.

\paragraph{The cheapest correct refresh.} When the antecedent does change materially, the refresh is cheap. Swapping $d{\in}\{0,1,2,3\}$ predecessor frames for distractors, a rank-$32$ patch \emph{update} tracks full rebuild ($\eta{=}0.81$--$0.91$ on GQA, $0.94$--$0.98$ on deepstack, within $0.02$--$0.06$ of full re-prefill) at $\sim\!m/F$ of its bytes, so a set-change is absorbed by a memory-axis update rather than a recompute. Depth again decides the recompute alternative: re-prefilling the deep three-quarters holds ($\ge\!0.90$) while a shallow-only refresh is unreliable (down to $0.37$ at $d{=}3$), and a \emph{shared} basis lags both. The reuse envelope is therefore bounded to the recurring/related-antecedent regime, with a rank-$32$ patch update as the cheapest correct response to a changed set; arbitrary-context serving is handed to re-prefill.

\section{Context management as reversible state edits}
\label{app:ctx}

Because the patch is additive and the position-free canonical is already what we store, conditioning is a \emph{switch} with an asymmetric cost. Reverting to the context-clean view is free: overwrite the chunk's cache entries with the stored canonical $\kvo$, a copy with no arithmetic at all. Re-applying conditioning is a single rank-$m$ add, $\kvo+U_mV_m^{\!\top}$. Both directions are exact and reversible. The body measures one consequence (reversible eviction, \S\ref{sec:window}); two further serving patterns follow, which we flag as design-space openings rather than measured results.

\paragraph{A near-free disposability test.} To ask whether an antecedent is still load-bearing for a pending query, toggle its patch \emph{off} and decode on the canonical: if the answer is unchanged the context was not needed and can be dropped, a check that costs one decode rather than a recompute. This is the cheap, query-specific counterpart to an orchestrator's semantic-liveness guess, and it sidesteps the diffuse-token result (\S\ref{sec:shape}) that rules out attention-magnitude importance heuristics.

\paragraph{Copy-on-write speculative forks.} Because the canonical is shared, an agent can branch onto a context-clean fork while keeping the conditioned overlay separate (continuing useful work while a summarizer compresses the live context, or fanning out a search) and pay to re-condition only on the branch it commits, at the cost of a per-branch patch rather than a duplicated cache. Tree-search visual agents are the clearest instance: ZoomEye~\citep{zoomeye2024} and V$^*$~\citep{wu2024vstar} explore many zoom/crop branches over one shared image and commit the highest-confidence path; PixelCraft~\citep{pixelcraft2025} maintains an image memory so its planner can revisit earlier visual steps, today paid as a full re-encode per branch.

\paragraph{Reuse-aware placement and scheduling.} Because reorder is free over the permutation orbit (\S\ref{sec:window}), the chunks a window holds are effectively a \emph{set}, and their arrangement in the deque is a free variable rather than a consequence of arrival order. This turns context assembly into a scheduling problem with two coupled decisions: which chunks to admit under a fixed memory budget, and in what order to place them so that cached patches stay valid (a chunk lands behind an antecedent it has already been conditioned on, so its stored patch is reused rather than reformed) and the per-step conditioning cost is minimized. A prefix cache cannot pose this question, since placement there is dictated by position and any reorder is a miss; the position-free store makes chunk placement an optimization target in its own right, trading patch-forming cost against memory and reuse hit-rate. Characterizing this objective and the policies that optimize it, online as a window slides and offline over a known access trace, is left to future work.

\paragraph{Two boundaries.} First, the scheme is \emph{forward}-lossless (a chunk with no future direct read is free to drop), not a retroactive edit: an evicted chunk's already-absorbed influence on the surviving chunks cannot be \emph{exactly} inverted, because cross-chunk conditioning is not an invertible linear operator. We measure that influence to be small (survivors are near-lossless to keep as-is on GQA and MLA; \S\ref{sec:window}). Second, re-instating a chunk that is itself an antecedent for other cached chunks may \emph{cascade}, triggering conditioning patches on its in-cache dependents, so the patch composes into a dependency graph of re-materialization rather than a single local edit. Characterizing that cascade, and the agentic workloads where reversible context management pays, is left to future work.

\end{document}